# Refining microstructures in additively manufactured Al/Cu gradients through $TiB_2$ inclusions

Michael J. Abere,[1]* Hyein Choi,[1] Levi Van Bastian,[1] Luis Jauregui,[1] Tomas F. Babuska,[1] Mark. A Rodriguez,[1] Frank W. DelRio,[1] Shaun R. Whetten,[1] and Andrew B. Kustas[1]

[1]Sandia National Laboratories, Albuquerque, NM USA 87123

*Corresponding author. *E-mail address:* mjabere@sandia.gov (M.J. Abere)

**Abstract**

The additive manufacture of compositionally graded Al/Cu parts by laser engineered net shaping (LENS) is demonstrated. The use of a blue light build laser enabled deposition on a Cu substrate. The thermal gradient and rapid solidification inherent to selective laser melting enabled mass transport of Cu up to 4 mm away from a Cu substrate through a pure Al deposition, providing a means of producing gradients with finer step sizes than the printed layer thicknesses. Printing graded structures with pure Al, however, was prevented by the growth of $Al_2Cu_3$ dendrites and acicular grains amid a matrix of $Al_2Cu$. A combination of adding $TiB_2$ grain refining powder and actively varying print layer composition suppressed the dendritic growth mode and produced an equiaxed microstructure in a compositionally graded part. Material phase was characterized for crystal structure and nanoindentation hardness to enable a discussion of phase evolution in the rapidly solidifying melt pool of a LENS print.

## 1. Introduction

Additive manufacturing (AM) is a set of material deposition processes capable of creating complex form factors directly from computer modelled parts [1,2]. There has been particular interest in the AM of metals by means of selective laser melting (SLM) [3,4] due to the increasing needs for direct part production [1,5,6], light-weighting [7], and creation of part architectures that cannot be achieved by conventional methods, such as casting or machining [8,9]. As such, metal AM promises to revolutionize multiple industries by accelerating the design cycle [10,11]. The techniques enable integration of unique geometries and material functionalities [12-14] while simultaneously consolidating or eliminating traditional production assemblies [15,16]. The most common SLM technique for structural material applications is laser powder bed fusion [3-5]. AM via laser engineered net shaping (LENS) [4,17] has garnered recent attention due to its ability to grade material composition locally during deposition [11,18,19]. In contrast to functional grading in powder bed and non-laser-based techniques like wire arc direct energy deposition [20], LENS utilizes a concentric inert gas powder feed system sprayed directly at the focal point of the build laser. The powder feeds can each contain a unique metal or alloy such that compositional grading is achieved by varying the powder feed rates in a given layer.

One material system of interest for compositional grading is the Al/Cu system due to the resulting heterostructures, combining the corrosion resistance and lightweight properties of Al with the high thermal/electrical conductivity of Cu. The material system has had application across a diverse set of fields including power electronics [21], aerospace [22], communications [23], and high energy density physics [24]. The ability to functionally grade the Al/Cu material system has the potential to produce composites that exhibit properties that surpass those of its constituents [25]. Producing graded Al/Cu structures with LENS has the advantages of producing such parts millimeters in height [11,18] with reported densities exceeding 99% of full [26]. AM also enables local microstructural control through tuning process parameters [27-29]. However, there exist specific challenges associated with grading the Al/Cu system. Namely, the intermetallic grains tend to grow in a preferentially aligned columnar mode epitaxially from seed crystals in the underlying layer that result in dendrites after laser melt and resolidification that prevent uniform composition within a given plane [30-33]. Eliminating this microstructure is further complicated by the large thermal gradients inherent in LENS that exceed those in other AM processes [34].

Process parameters have been optimized across numerous materials systems in attempts to reduce or eliminate columnar dendritic features in AM alloys [32,33]. Changing energy density has been explored for controlling growth mode [35,36] as well as powder flow rate [36-39]. Collectively, these studies found that a higher powder feed rate promoted a transition from columnar to equiaxed solidification during processing. The relationship with energy density is more complex as higher densities on a bulk part enable less kinetic undercooling for columnar growth [35], but in an AM process unmelted powder acts as nucleation sites for equiaxed grains more frequently at lower laser energy densities at rates sufficient to overwhelm bulk thermo-kinetic predictions of the growth mode [36,37]. Columnar to equiaxed transitions in AM alloys have also been promoted chemically through the addition of grain refiners or inoculants. Some common grain refining particle additions in Al are $La_2O_3$ [33,40], $TiB_2$ [41,42], $Al_2O_3$ [42-44], ZrH [42,45], and Ta [42]. In the case of Al alloys with $TiB_2$ particles, studies of cast parts have determined that the particles act as heterogeneous nucleation sites to seed new equiaxed grains ahead of the solidification front that interrupt growth of columnar structures [46,47]. This work combines the grain refining properties of $TiB_2$ inoculants in Al with the compositional grading capabilities of LENS to tailor the AM melt pool such that the growth of columnar dendrites at an Al/Cu interface is completely suppressed. Achieving an equiaxed grain growth mode enables the deposition of mm-scale compositionally graded Al/Cu parts.

## 2. Materials and Methods

Feedstock material included Cu, Al, and $TiB_2$ gas-atomized powders from American Elements, Los Angeles CA. The Al and Cu powders had a nominal 45 mm – 105 mm size distribution. The $TiB_2$ powder had a nominal size of 170 mesh. A Nuburu fiber laser operating at 450 nm was mounted to an optics port on a 3-axis Tormach CNC 770 (Tormach LLC., Waunakee WI) housed in a controlled atmosphere glovebox (MBraun, Inc., Strathan, NH). The choice of a 450 nm wavelength was based on the improved density compared to infrared lasers in welding applications of Cu [48]. Attempts to deposit Al onto Cu substrates with a 1064 nm infrared laser were unsuccessful in depositing discrete layers of material and damaged optics in the tool from backreflection when power was increased to drive melt at the interface. The atmosphere consisted of a continuously purging Ar gas to maintain an atmosphere of <50 ppm $O_2$ and <10 ppm $H_2O$. The powder feeding system allowed for independent control of each powder flowrate by combining a rotary feed wheel and carrier gas to deliver fluidized feedstock. Further details on the LENS AM tool can be found in Ref. [49].

Additively manufactured cylinders were deposited in the blue-light LENS system. Cylinders were deposited that were 5 mm in diameter and 4 mm tall with a 0.25 mm layer thickness. Al powders were mixed in a bottle that was then placed on a bottle tumbler for 30 minutes with 0-2 wt% $TiB_2$ to produce custom alloys for grain refinement studies. The 2 wt% upper bound was chosen to avoid excessive increase in the composite density, targeting a maximum increase of 1.3%. The alloyed Al and pure Cu powders were fed through the LENS system with a gas flow rate of 3 liters per minute. Laser power was held fixed at 600 W, and spot diameter was reduced to 1.2 mm by focusing the beam with a 100 mm focal length meniscus lens. Laser power was measured using a FL1100A-BB-65 fan cooled thermal power meter. The build plate temperature was preheated to 300˚C using a cartridge-based heat exchanger. The LENS toolpath was followed at a speed of 12 mm/s for the part perimeter followed by 15 mm/s for the infill. Samples were deposited as pure Al in direct contact with a Cu substrate, grain refined Al on a Cu substrate, and grain refined Al on top of different concentrations of Al/Cu mixed layers to characterize the resulting microstructure and compositional variation present at various interfacial conditions in a density graded structure.

Sample phase was characterized via X-ray diffraction and local phase information was determined through elemental mapping in micro-X-Ray fluorescence (μXRF). Sample microstructure was imaged both optically and in scanning electron microscopy (SEM) and the contrast was correlated to material phase via electron backscatter detection (EBSD). Chemical composition of microstructural features was characterized in energy dispersive X-ray spectroscopy (EDS).

Nanoindentation measurements were performed with a diamond Berkovich tip on a Hysitron TI-980 Triboindenter. Prior to Al/Cu measurements, the tip area function and load frame compliance were calibrated over the entire load range of the instrument with fused silica as the reference. For the Berkovich tip, the tip area function was defined via $A(h_c) = C_0 h_c^2 + C_1 h_c + C_2 h_c^{1/2} + C_3 h_c^{1/4} + C_4 h_c^{1/8}$, where $h_c$ is the contact depth and $C_0$ through $C_4$ are coefficients related to tip shape. Herein, $C_0$ was taken to be the ideal value for a Berkovich tip ($C_0$ = 24.5), and $C_1$ through $C_4$ were found through fits to the calibration data. The fit between the data and model validated the area function down to $h_c$ of 20 nm. For the measurements on Al/Cu parts, ultrahigh-speed property maps (XPM) were generated over 60 mm × 60 mm areas to evaluate spatial heterogeneities in the properties at Al/Cu interfaces with various $TiB_2$ concentrations. The indent spacing was 3 mm; the spacing was large enough to prevent interactions between neighboring indents [50] but small enough to facilitate high-resolution maps of the microstructural features. The indents were performed in load control to a force $F$ of 1000 mN; this force translated to $h_c$ of ≈30 to 170 nm, which were large enough to diminish surface effects but small enough to minimize substrate effects. The unloading segment of each $F$-$h_c$ curve was analyzed with the Oliver-Pharr method [51] to determine reduced modulus $E_r$ and hardness $H$. $E_r$ was defined by $E_r = (S/\sqrt{A})(\sqrt{\pi}/2)$, where $S$ is the stiffness of the upper portion of the unloading curve. $H$ was defined by $H = F_{max}/A$, where $F_{max}$ is the maximum force. The resulting $E_r$ and $H$ were used to assess properties for different microstructural features.

## 3. Results and Discussion

### 3.1. Characterization of Microstructure Refinement

Al was deposited via LENS on Cu substrates with varying concentrations of $TiB_2$ mixed within the powder feedstock. The additively manufactured Al with 0 wt%, 1 wt%, and 2 wt% $TiB_2$ on Cu was cross-sectioned, polished, and characterized optically. The resulting effects of the $TiB_2$ additions on the morphology of the Al/Cu interface, with respect to unmodified Al/Cu, is shown in the optical micrographs in Figure 1(a-c). The pure Al deposition in Figure 1(a) forms a dendritic structure that extends 500 µm high along the build (vertical) axis followed by 200 µm of acicular grains (total height of 700 µm from the interface). Faceted grains are optically visible further away from the interface. The addition of 1% $TiB_2$ to the Al powder reduces the height of the dendritic structure at the Al/Cu interface to 400 µm in Figure 1(b). The acicular grains now extend 600 µm beyond the interface. The observed trend continues when increasing the $TiB_2$ concentration in the Al to 2 wt% as seen in the Figure 1(c) optical image, where the dendrites only extend 200 µm beyond the Al/Cu interface with the acicular to faceted grain transition occurring after 400 µm of total deposition. While the 98 wt% Al, 2 wt% $TiB_2$ alloy deposition managed to reduce the $Al_2Cu$ intermetallic grain growth, the irregularity of the dendrite and acicular grains would prevent continuous grading throughout the part.

Utilizing the in-situ grading capabilities of LENS, a buffer layer of 67 wt% Cu and 33 wt% 98 wt% Al, 2 wt% $TiB_2$ was deposited on the Cu before building the remaining layers out of 98 wt% Al, 2 wt% $TiB_2$. The 67/33 wt% (47/53 at%) powder mixture has a concentration between the AlCu and $Al_2Cu$ intermetallics in the Al/Cu materials system. An optical micrograph of the resulting interface is shown in Figure 1(d). This study kept laser and tool path parameters constant to isolate the effects of the buffer layer composition while minimizing changes to thermal gradients. While the deposited material has an increase in void population without optimizing the print parameters, the dendritic and acicular grains are removed from the microstructure. In short, compositional grading in LENS enables equiaxed grain growth directly at an Al/Cu interface.

The dendrite to equiaxed transition in printed microstructures was further characterized for phase and structure. The pure Al deposition was imaged in SEM with a backscatter detector and local phase information was garnered from EBSD coupled to EDS. An SEM image of the Al/Cu interface and optical image maps (OIM) from selected areas of the various microstructures are provided in Figure 2. Within the OIM, the intermetallic phases are indexed by their crystal symmetries and lattice parameters (hexagonal $Al_2Cu_3$ and tetragonal $Al_2Cu$). However, the FCC Al and Cu with a 12 wt% difference in lattice parameter -- plus any disordered solid solution thereof -- are not differentiable from their Kikuchi patterns alone. When coupled with the Z contrast from the backscatter detector in the SEM image of Figure 2(a) and EDS data (see Supplementary Material), it is apparent that the cubic phase in the substrate shown in region 1 of Figure 2 is Cu while the cubic phase that appears above the faceted $Al_2Cu$ grains in region 3 of Figure 2 is Al. Given this information on interpreting the color map in the OIM, the SEM contrast in Figure 2(a) can be correlated to phase with each elemental metal and intermetallic having uniform Z contrast.

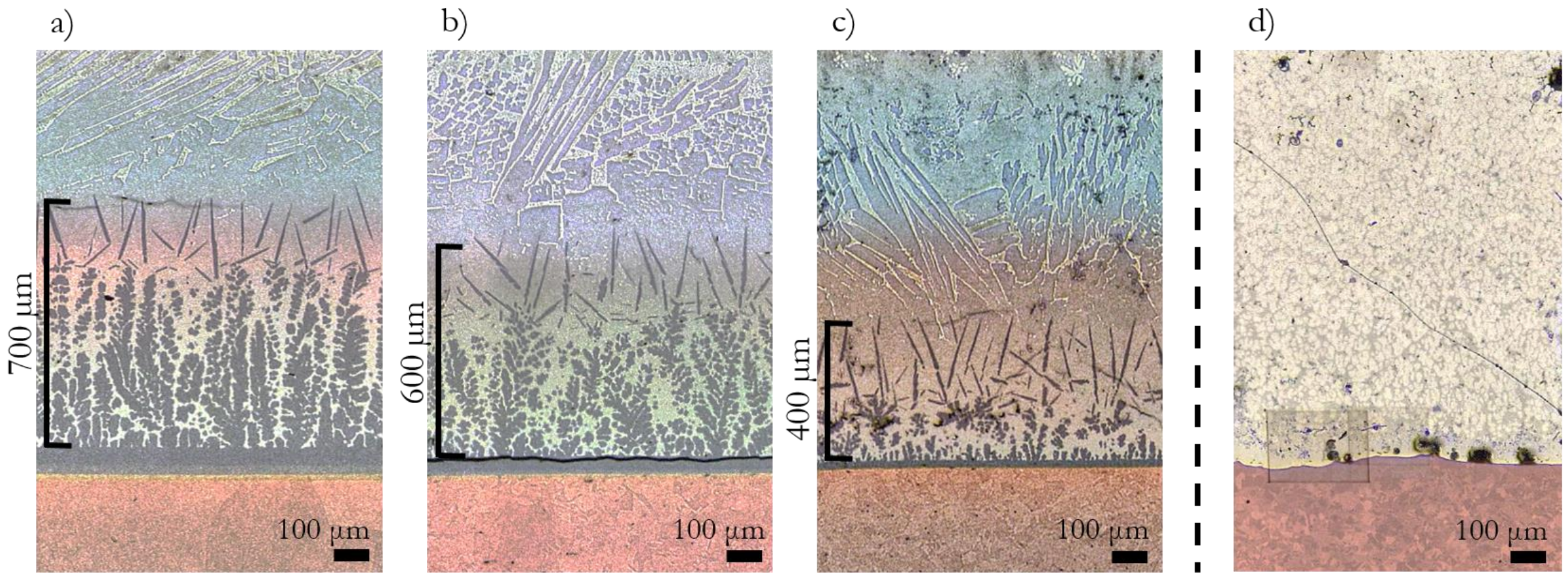

**Fig. 1.** Optical microscopy of resulting morphology at Al/Cu interface. (*a*) pure Al, (*b*) addition of 1 wt% $TiB_2$ to Al powder, (*c*) addition of 2 wt% $TiB_2$ to Al powder, and (*d*) contains a 67/33 wt% powder mixture of Cu/(98 wt% Al, 2 wt% $TiB_2$) before printing 98 wt% Al, 2 wt% $TiB_2$. With greater additions of $TiB_2$ (*a*), (*b*), (*c*) a decrease in columnar grain height is observed with complete suppression in (d).

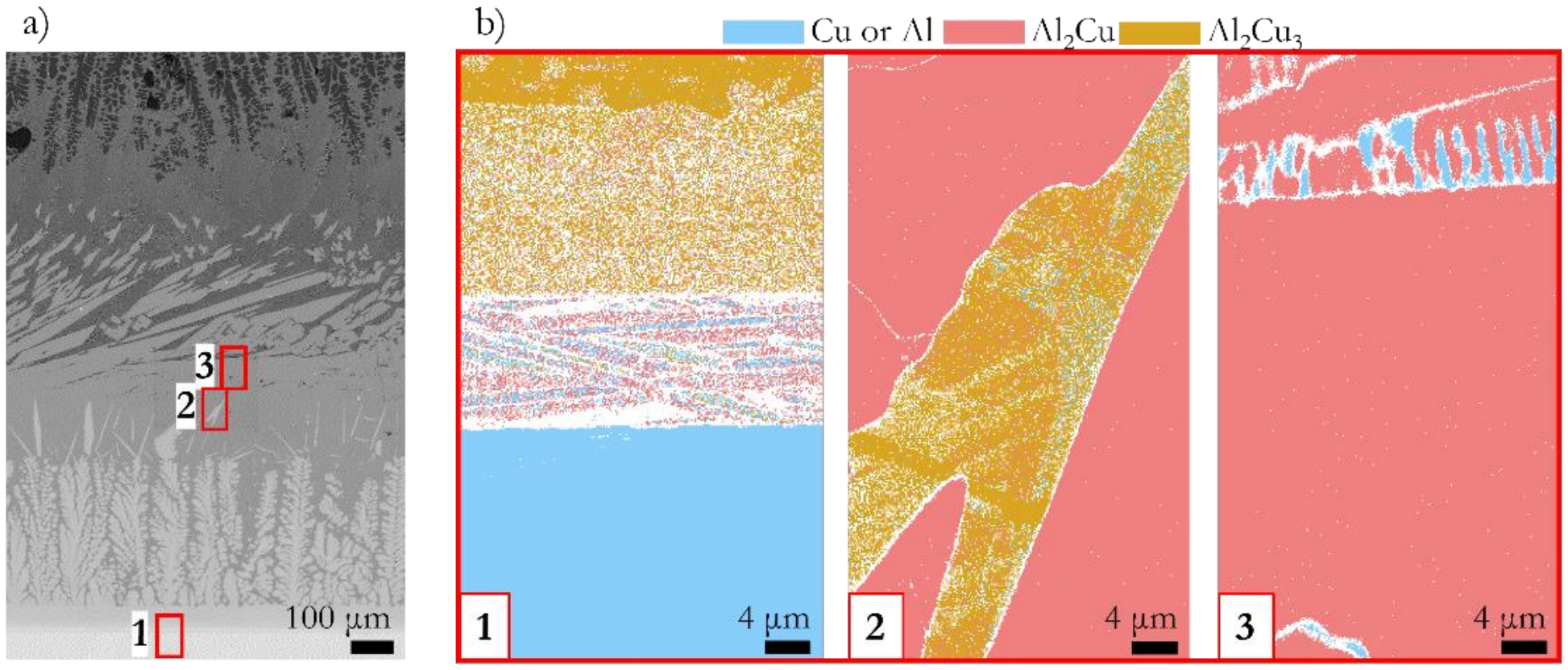

**Fig. 2.** SEM image of the deposition of pure Al on Cu (*a*) depicts larger area with rectangular boxes correlating to numbers 1 through 3. EBSD patterns taken from the inset locations of each numbered region can be seen in (*b*) with a legend of chemical compounds above.

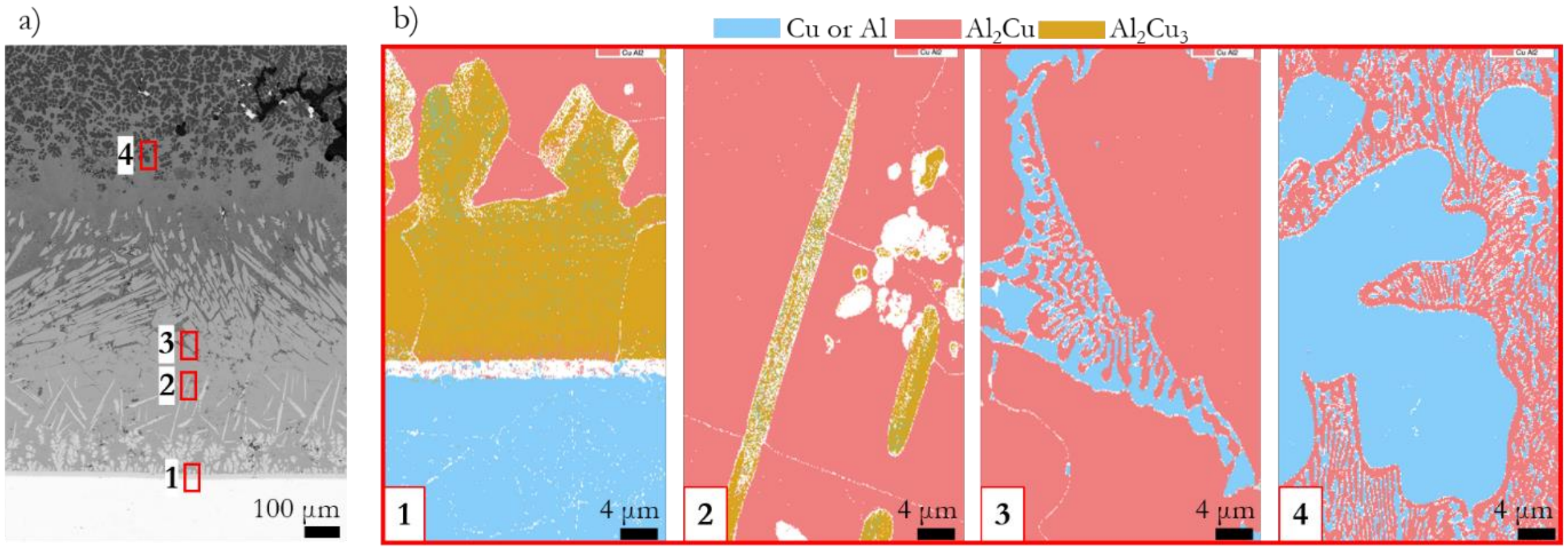

**Fig. 3.** SEM image of the Cu interface after deposition of 2 wt% $TiB_2$. (*a*) depicts larger area with rectangular boxes correlating to numbers 1 through 4. EBSD patterns taken from the inset locations of each numbered region can be seen in (*b*) with a legend of chemical compounds above.

The deposition of pure Al on Cu produces two continuous layers at the interface seen in the OIM from region 1 of Figure 2 before seeding the observed dendritic growth. The first ~10 µm of printed material forms a layer that does not produce well defined Kikuchi bands upon diffraction (see Supplementary Material), which suggests a highly strained lattice, a partial loss of crystalline order. The collected EBSD signal suggested a mixture of cubic and tetragonal symmetries while the Z contrast is intermediate to Cu and $Al_2Cu_3$. Atop the highly strained interlayer is a continuous ~15 µm of $Al_2Cu_3$. The root of an $Al_2Cu_3$ dendrite growing from this continuous seed layer can be seen at the top of the OIM of region 1 in Figure 2. The dendrites and acicular grains are composed of $Al_2Cu_3$ and grow into an $Al_2Cu$ matrix. Additional intermetallic phases are notably absent. The $Al_2Cu$ grains also grow with a columnar mode into an Al-rich matrix composed of a two-phase composition. The OIM from region 3 of Figure 2 shows that at the edge of $Al_2Cu$ growth, a two-phase hypereutectic mixture of $Al_2Cu$ and Al are observed, consistent with the equilibrium phase diagram.

The addition of $TiB_2$ to the printed Al did not change the phase of the columnar dendrite and acyclic grains. An SEM image with backscattered electron contrast of printed Al and 2 wt% $TiB_2$ on Cu with EBSD from selected areas is shown in Figure 3. The root of the dendrite and tip of an acyclic grain within the OIMs from inset regions 1 and 2 in Figure 3 are both composed of $Al_2Cu_3$ while having uniform Z-contrast in the backscatter SEM image. Similar to the pure Al deposition, the $Al_2Cu$ matrix containing the $Al_2Cu_3$ grains exhibits columnar growth into a two-phase hypereutectic matrix of $Al_2Cu$ and Al (see inset 3 in Figure 3). In contrast, addition of $TiB_2$ to Al changed the height within the print at which hypoeutectic α-Al nucleated. For the 2 wt% deposition, this microstructure is visible within inset 4 of Figure 3 at 800 µm above the Cu interface. The same is not seen until 1100 µm for pure Al deposition in Figure 2.

An XRD pattern of the dendrite-free 98 wt% Al, 2 wt% $TiB_2$ deposition is shown in Figure 4(a). While the elemental Al and $Al_2Cu$ intermetallic are still present in the deposited material, there are no longer peaks for a hexagonal $Al_2Cu_3$. Color maps of µXRF signal for Cu, Al, and Ti from the cross-sectioned sample are provided in Figures 4(b), (c), and (d), respectively. Qualitatively, the µXRF shows that the printed material is compositionally graded between Al and Cu. The Ti signal is strongest either near the Al/Cu interface where relative Cu concentration is the highest or at the top of the print where the composition is most Al rich. There is a noted deficit of the grain refiner between these two compositional zones. This observation is consistent with grains nucleating from the TiB2 surfaces near the interface to selectively sequester the element from the melt [52] where grain refinement is most pronounced. The Al/Cu interface was more closely characterized with SEM, EDS, and EBSD in Figure 5. Directly at the Cu interface is a thin, 1 µm layer of $Al_2Cu_3$. The phase did not appear in the XRD pattern because the X-ray beam was placed near the interface but did not cross the substrate boundary. Above the thin $Al_2Cu_3$ layer is 2 µm of θ-$Al_2Cu$ followed by a two-phase region of θ-$Al_2Cu$ and α-Al. The two-phase region starts with submicron diameter, equiaxed α-Al precipitates in an θ-$Al_2Cu$ matrix, consistent with the θ+α hypereutectic microstructure in the Al/Cu binary. As the local Al concentration increases, the precipitate diameter also increases until its concentration reaches the invariant point, and the lamellar eutectic microstructure is observed. The lamellae are only 580 ± 15 nm wide, which make them ideal for micron-scale compositional grading in the Al/Cu materials system. Further increases in the local Al concentration led to the nucleation of larger, 3 ± 1 µm diameter Al precipitates within the eutectic lamellar structure consistent with an α+θ hypoeutectic microstructure in the Al/Cu binary. These larger precipitates represent a drastic increase in compositional gradient step size compared to the θ+α hypereutectic and pure eutectic microstructures. The α+θ hypoeutectic microstructure continues throughout the print to a total height of >4 mm with the α-phase precipitates increasing in diameter to 15 ± 5 µm, 4 mm away from the interface, as seen in Figure 6 alongside the resulting compositional gradient from the AM print. The Al concentration gradient exhibits exponential decay between 67% and 99% Al with step functions between pure 0% Al, 40% Al, and 67% Al. Control over the gradient shape and methods to grade over compositions more Cu rich than the $Al_2Cu$ phase will be a focus of future work.

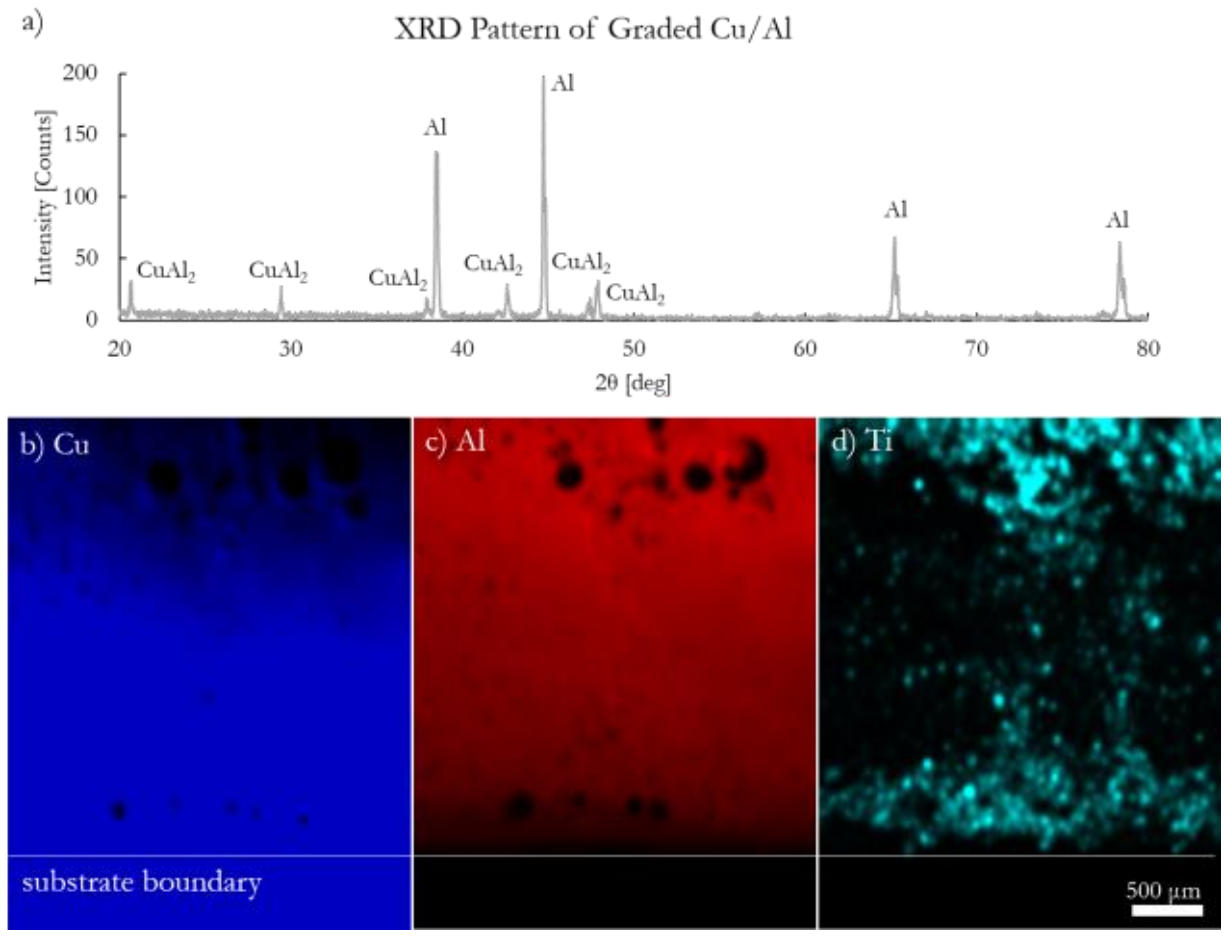

**Fig. 4.** (*a*) XRD pattern of the dendrite-free 98 wt% Al, 2 wt% $TiB_2$ deposition. The XRD pattern shows that the material lacks peaks of monoclinic AlCu, but Al and $Al_2Cu$ intermetallic still present. (*b*), (*c*), (*d*) provide color maps of µXRF of Cu, Al, and Ti respectively. The printed material appears compositionally graded between Al and Cu.

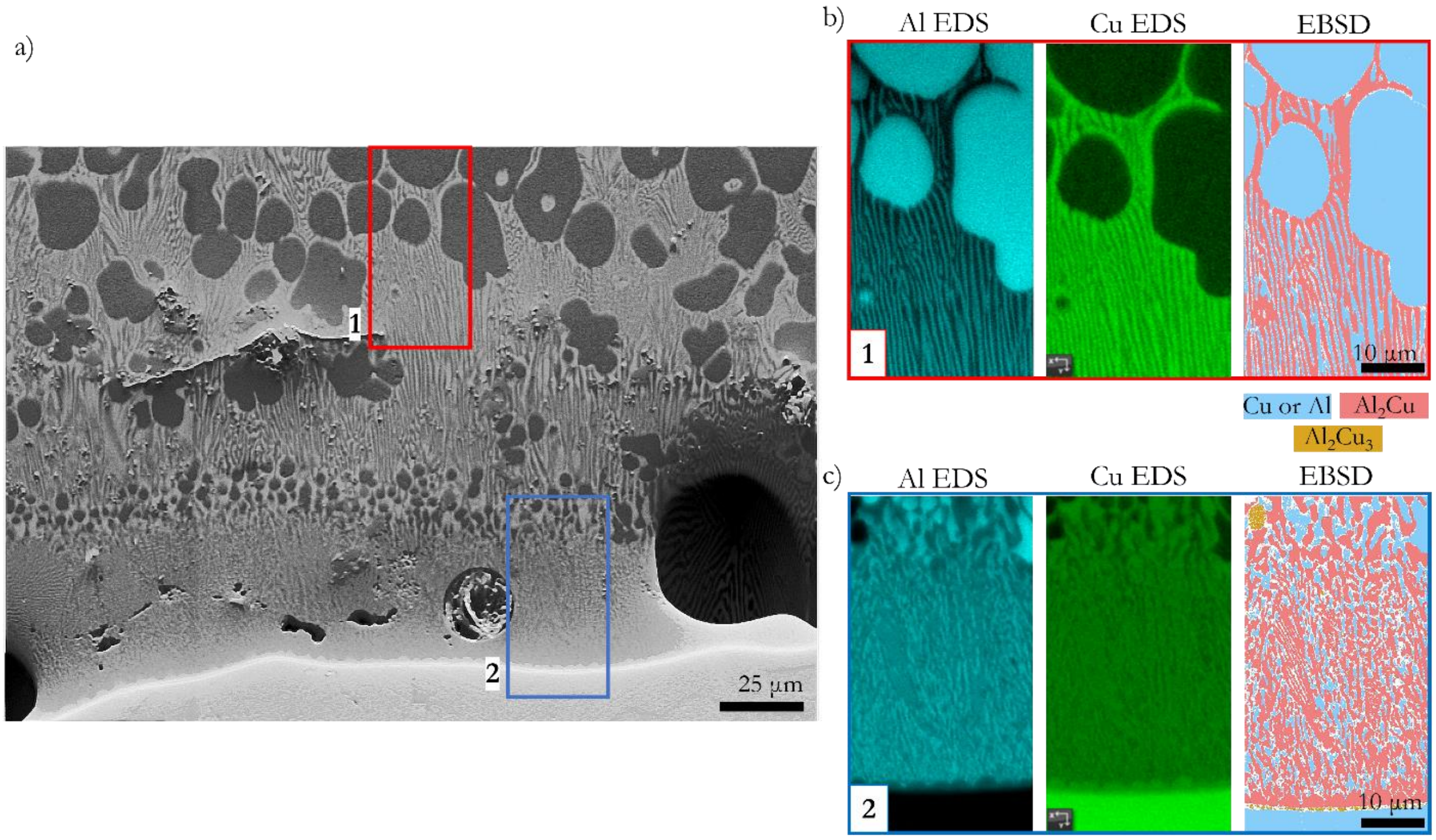

**Fig. 5.** SEM, EDS, and EBSD characterization of an Al/Cu interface after printing a 67/33 wt% Cu/(98 wt% Al, 2 wt% $TiB_2$) buffer layer on Cu followed by 98 wt% Al, 2 wt% $TiB_2$. At the Cu interface, a thin $Al_2Cu_3$ layer is followed by $Al_2Cu$ then a two-phase region of θ-$Al_2Cu$ and α-Al are seen. (*b*) and (*c*) are enlarged insets of the red and blue boxed regions with a legend for EBSD to the right.

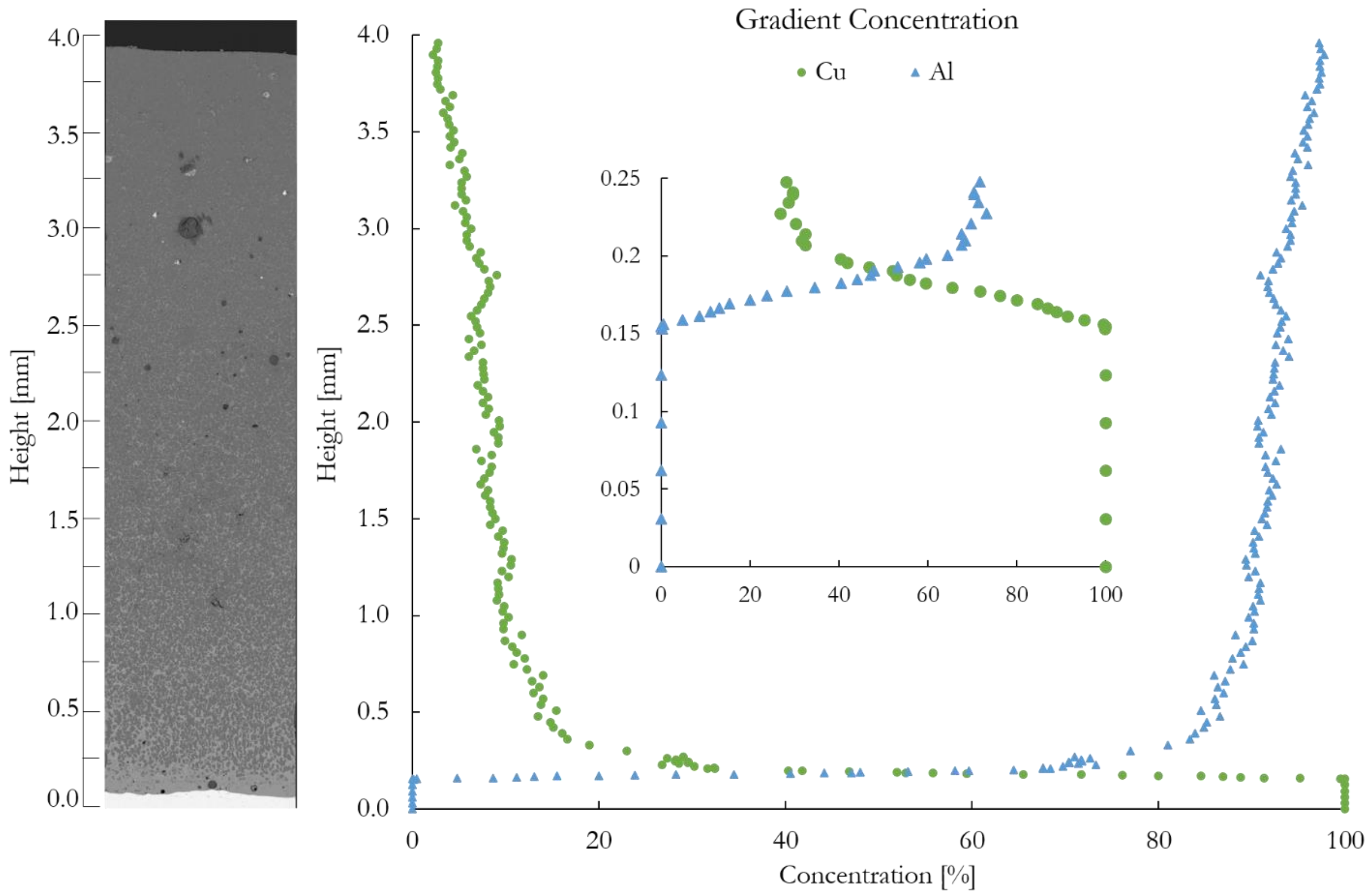

**Fig. 6.** SEM of complete print height of a 67/33 wt% Cu/(98 wt% Al, 2 wt% TiB2) buffer layer on Cu followed by layers of 98 wt% Al, 2 wt% TiB2 and the resulting compositional gradient plot. Drastic increases in compositional gradient step sizes occur as the larger Al precipitates dominate much of the concentration after about 0.5 mm of deposition. The build plate is pure Cu with a 1 µm layer of Al2Cu3 that forms at the print interface followed by two-phase regions of θ-Al2Cu and θ+α. The graph inset shows the compositional gradient at the Al/Cu interface.

### 3.2. Nanoindentation Hardness Properties

Results of nanoindentation testing in selected areas of the Al/Cu heterostructures are shown in Figures 7 and 8. Optical images, corresponding nanoindentation hardness maps, and histograms from each selected area are shown for both the pure Al (Figure 7) and the sequential print of a 67/33 wt% Cu/(98 wt% Al, 2 wt% $TiB_2$) buffer layer on Cu followed by 98 wt% Al, 2 wt% $TiB_2$ (Figure 8). Equivalent nanoindentation maps of the prints from Figure 1(b) and (c) are provided in the Supplementary Material. The nanoindentation hardness at the interface between the Cu substrate and pure Al deposition, shown in Figure 7(a) shows three distinct regions. The Cu substrate has an average hardness of 0.95 GPa, which is consistent with bulk values for Cu [53]. The interlayer between the Cu substrate and $Al_2Cu_3$ intermetallic that did not index well in EBSD has an average hardness of 2.9 GPa. The value falls below reported values of intermetallics $Al_2Cu$, AlCu, $Al_3Cu_4$, $Al_2Cu_3$, and $AlCu_4$ [54,55] while a solid solution rule of mixtures between Cu and even a fine-grained Al metal would predict a hardness below the elemental value for Cu [56]. Instead, the observed hardness of that layer in Figure 7(a) is equivalent to the homogeneous supersaturated Al α-FCC Cu phase formed from co-sputtering composite targets [57]. Values for nanoindentation hardness of the $Al_2Cu_3$ seed crystal phase vary between 7.4 GPa (cold roll-bonded) [58] and 14.9 GPa (furnace welding) [54], and the measured values of that phase in Figure 7(a) of 7.5 GPa falls within that range. The full width half maximum of 1.9 GPa locally throughout the $Al_2Cu_3$ on a single sample as well as the broad range of reported values suggests the hardness of the intermetallic is highly sensitive to variations in micro/nanostructure.

The nanoindentation hardness of $Al_2Cu_3$ dendrites within an $Al_2Cu$ matrix is shown in Figure 7(b). The intermetallic hardness of the seed layer at 7.5 GPa is still visible within the histogram for the intermetallic grains. However, the dendrite in the upper right corner of the interrogated region has a hardness exceeding 12 GPa, which is a value still within the range of previously reported values for $Al_2Cu_3$ but further demonstrates the local micro/nanostructural sensitivity of this phase's hardness. The $Al_2Cu$ phase appears with a hardness of 5.6 GPa, which is within the range of reported values of 5.0 GPa (hot pressing), 5.4 GPa (cold roll-bonded) [58], and 10.9 GPa (furnace welding) [54]. The acyclic $Al_2Cu_3$ grains in Figure 7(c) retain the same hardness as their seed layer and dendrite forms. The α-Al grains that nucleate within the $Al_2Cu$ matrix in **Fig. 8.** Nanoindentation hardness measurements of a 67/33 wt% Cu/(98 wt% Al, 2 wt% TiB2) buffer layer on Cu followed by layers of 98 wt% Al, 2 wt% TiB2. Local hardness is characterized at different regions of the sample including (*a*) the Al/Cu interface and (*b*) in the Al-rich topmost area.

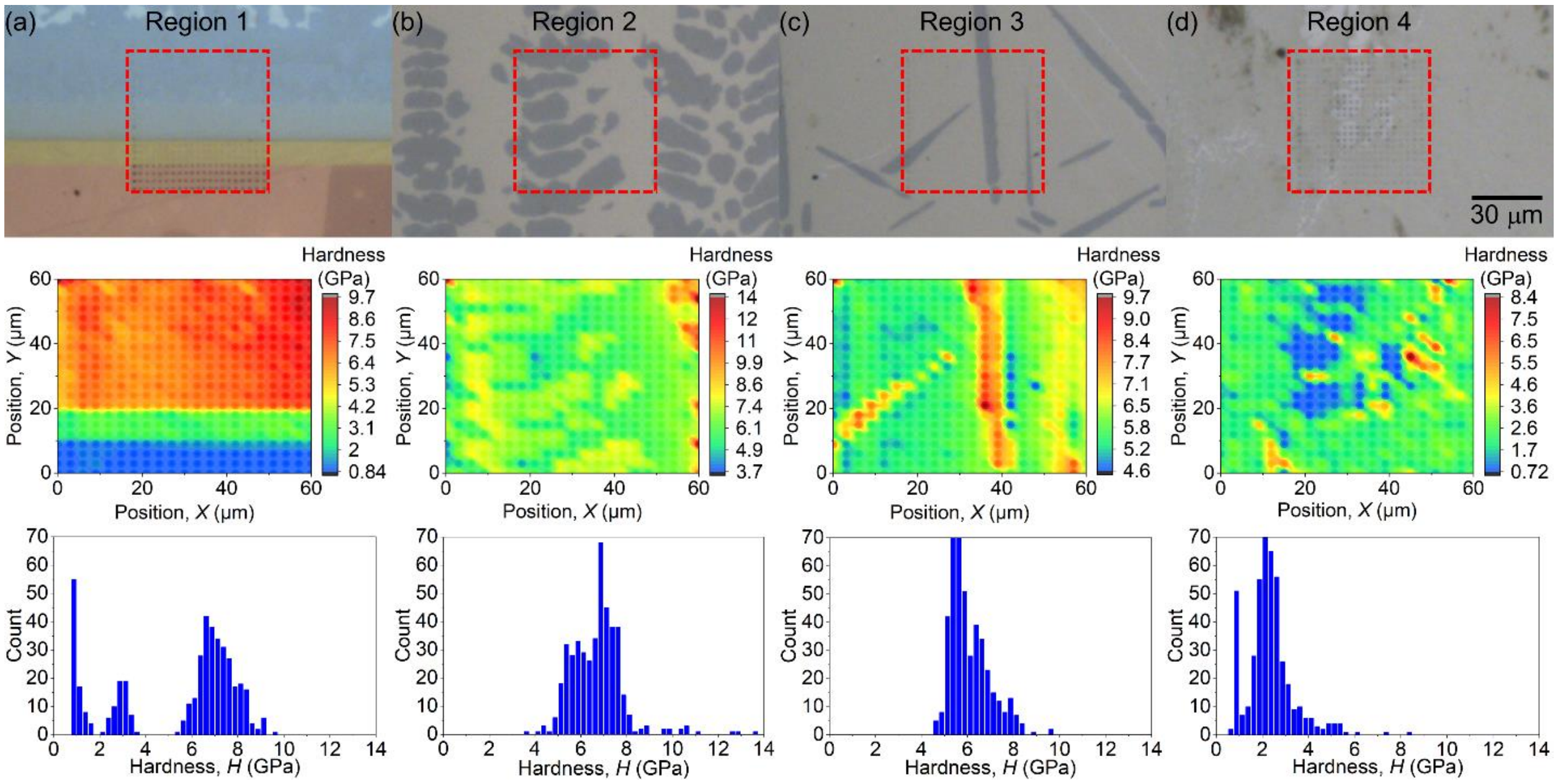

**Fig. 7.** Nanoindentation hardness measurements of a pure Al print on a Cu substrate showing local variation in different regions of the sample including (*a*) the Al/Cu interface, (*b*) the dendritic zone, (*c*) near acyclic grains, and (*d*) the Al-rich topmost area.

The compositionally graded interface arising from printing a 67/33 wt% Cu/(98 wt% Al, 2 wt% $TiB_2$) buffer layer on Cu followed by 98 wt% Al, 2 wt% $TiB_2$ is shown in Figure 8(a). Within the interfacial zone, the intermetallic grain size is smaller than the nanoindenter tip size, which leads to spatially average hardness measurements for each data point. While the relative phase fraction of $Al_2Cu$ and Al varies in the build direction, there is no apparent gradient in the hardness. The measured hardness of 6.1 GPa being higher than either of the two constituent phases is attributed to Hall-Petch strengthening from their nanograin structure. Interestingly, when the overall composition crosses the eutectic point such that the lamellae are Al-rich in Figure 8(b), the measured hardness resembles that of an intermediate value between the Cu-supersaturated α-Al and the $Al_2Cu$ intermetallic on the microscale.

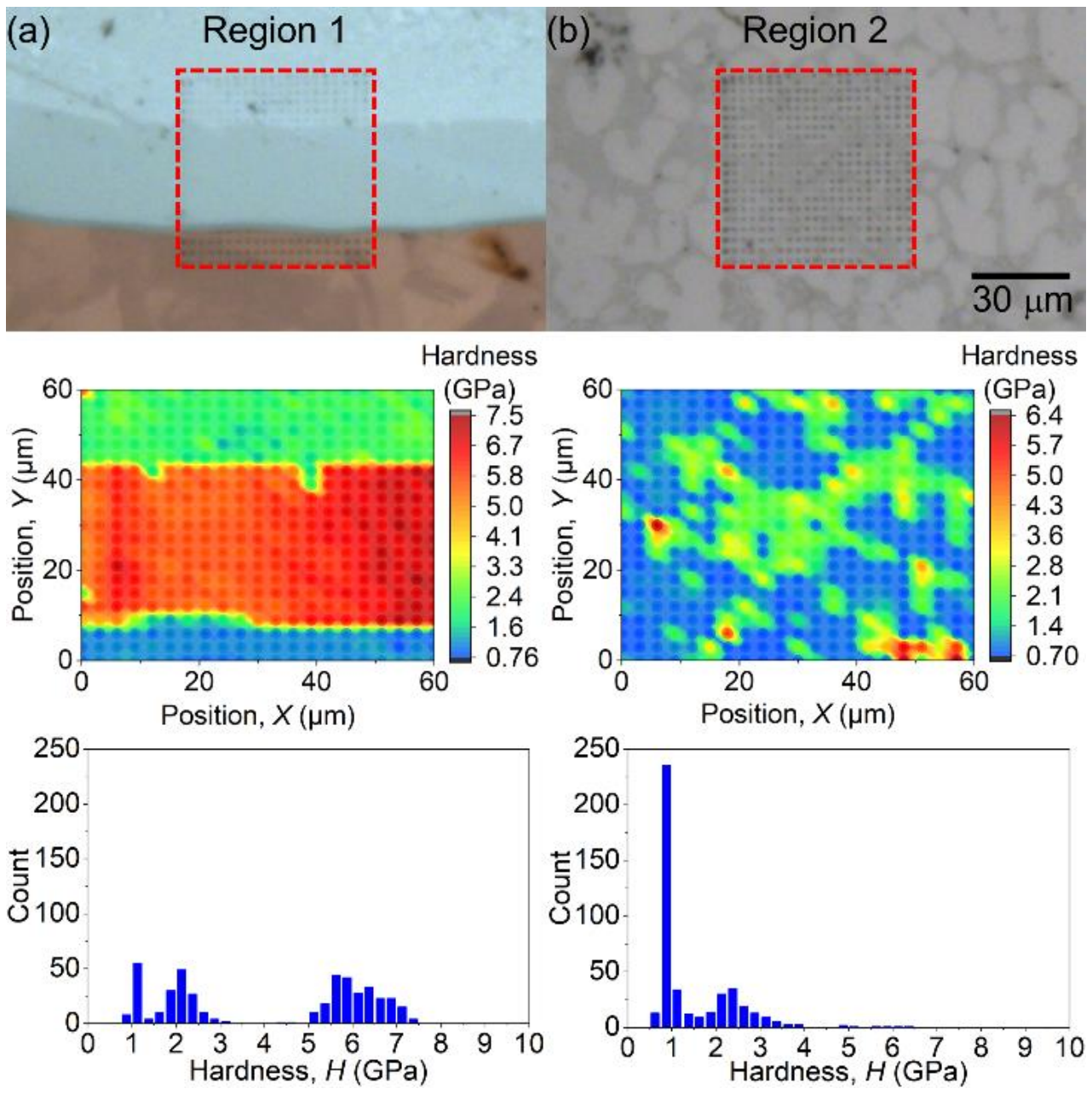

Figure 7(d) have hardness near 1 GPa, which is well above that of bulk Al [56] and instead matches that of α-Al with a supersaturated 4.5 wt% Cu in solid solution [54].

### 3.3 Discussion of Microstructural Refinement Mechanisms

The AM of dense pure Cu or printing on Cu substrates with conventional 1060 nm lasers has long proven difficult [59] without the use of additives [60] or volume reducing post process techniques [61]. The printability on Cu with a 450 nm blue laser can be attributed directly to the metal's optical coupling. The 15.4X higher solid absorptivity in Cu for 450 nm light over 1064 nm light [62] enables laser heating at substantially lower upstream powers, which reduces the intensity of light backreflected into the print head and allows delivery of sufficient intensity to melt Al powder on a Cu substrate. When considering laser melting of the substrate to form the interfacial weld, the magnitude of the discontinuity in optical reflectivity across the solid/liquid phase boundary of Cu becomes important. A calculation of the change in reflectivity for Cu between its liquid and solid phases showed that the difference in reflectivity between solid and liquid Cu shifted from 10.7X at a 1060 nm wavelength to a 1.9X difference at the blue laser's 450 nm wavelength [62,63]. This >5X narrowing of the differential absorptivity reduces the likelihood of superheating the melt pool onto the liquid/vapor isotherm for a given build laser output intensity. The importance of reducing melt pool temperatures in any AM process is that the recoil pressure within the AM melt pool is exponentially dependent upon the liquid temperature [64] and drives a depression within the liquid akin to a keyhole cavity [65]. The keyhole cavity collapse is a key source for porosity in AM parts [65-67] that cannot be mitigated through modified tool pathing [66].

Building an understanding of the suppression of columnar dendrites at the Al/Cu interface in a LENS printed part first requires a description of the active mechanisms capable of transporting Cu 4 mm away from the interface in the form of large intermetallic grains after pure Al was deposited onto Cu. That magnitude of Cu atom migration is >2000X greater than if the heterostructure was fabricated with an explosively welded interface [68]. Interestingly, the five orders of magnitude greater than ambient solid/solid interdiffusion transport during shock unloading in explosive welding of Al and Cu [68] is also observed for desorbed solid Cu atoms in liquid Al [69]. The key difference in the LENS printed parts is that the Cu source is the partial melting of the substrate and each subsequent layer in the build. Therefore, the movement of Cu atoms throughout the part is attributed to the combination of Marangoni effects and recoil pressure driving a turbulent (fluid velocities on the order of a m/s) co-melt of the deposited powder and underlying Cu rich layer [70].

The incorporation of the Cu mixed into the melt of each deposited layer into the observed $Al_2Cu$ intermetallic over other thermodynamically stable phases ($Al_4Cu_9$, $Al_3Cu_4$, and AlCu [71]) upon solidification involves a kinetically constrained nucleation and growth process. Within the rapidly solidifying AM melt pool, Arrhenius growth kinetics of each intermetallic favor the $Al_2Cu$ due to their lower activation energy and more exothermic effective heat of formation than the other competitive phases [72]. The preference for $Al_2Cu$ is further exacerbated by the abnormally high vacancy concentration (~1% [73] or $10^4$ above room temperature equilibrium [74]) inherent to rapidly quenched laser melts. Given that the physical mechanism for Cu diffusion into Al is vacancy site hopping [75] with net migration governed by the Kirkendall effect [76], transport of Cu atoms to more thermodynamically stable configurations is not unexpected. Annihilation of the far from equilibrium vacancy point defect concentration in Al commonly occurs via the formation of Guinier-Preston zones [77], which are the known nucleation sites of the $Al_2Cu$ intermetallic, specifically [78]. Once nucleated, the enhanced transport in the vacancy rich Al can also promote grain coarsening of the $Al_2Cu$.

The same competitive growth argument cannot explain the presence of $Al_2Cu_3$ as it has a less negative effective heat of formation than either AlCu or $Al_3Cu_4$ and empirically loses the competitive growth process to some combination of $Al_2Cu$, AlCu, $AlCu_3$, $Al_3Cu_4$, and $Al_4Cu_9$ in diffusion bonded [72], cold rolled [79], or welded [80] Al/Cu laminates. However, $Al_2Cu_3$ can grow at a liquid Al on solid Cu interface albeit in combination with $Al_2Cu$, AlCu, $AlCu_3$, and $Al_4Cu_9$ and much slower cooling rates (7 x $10^{-2}$ K/s) than occur in LENS [81]. At the rapid quench rates present in LENS and laser welding ($10^3$-$10^5$ K/s) [82], Zhang *et.al.* [83] also observed the isolated growth of $Al_2Cu_3$ followed by $Al_2Cu$ at an Al/Cu interface. Based on classical kinetic theory, wherein the thickness of a developed layer, $y$, that solidifies in time, $t$, at a given temperature, $T$, can be described by the equation [84]:

$$y = k_0 \exp\left(-\frac{Q}{RT}\right) t^n \qquad (1)$$

where $n$ is the time exponent, $Q$ is the activation energy, $R$ is the gas constant, $k_0$ is the frequency factor, and $k_0 \exp\left(-\frac{Q}{RT}\right)$ [86] is an Arrhenius expression for the intermetallic layer growth rate. Refinements of these kinetic constants from Guo *et. al.* [72] by Bedjaoui *et. al.* [54] by broadening the temperature range of mass transport measurements revealed that the $Al_2Cu_3$ intermetallic has a lower $Q$ (119 kJ/mol at.) than the $Al_4Cu_9$ phase (243.1 kJ/mol at.) but a smaller $n$ (0.41 for $Al_2Cu_3$ vs 0.47 for $Al_4Cu_9$). Such a relationship in kinetic parameters is consistent with the empirical observation that the $Al_2Cu_3$ phase forms under rapid quenching conditions with solidification nucleating near the liquidus temperature while the $Al_4Cu_9$ phase dominates in solid-state diffusional welds held at temperature for minutes. Bedjaoui *et. al.* [54] also calculated a similar relationship in which the $Q$ for the phase mixture of AlCu + $Al_3Cu_4$ (141.1 kJ/mol at.) is also larger than that of $Al_2Cu_3$, which is consistent with its growth in quenched laser melt pools but an $n$ of 0.40 does not universalize these kinetic constants for predicting intermetallic phases in all Al/Cu joining techniques where AlCu and $Al_3Cu_4$ are observed in the absence of $Al_2Cu_3$.

The addition of $TiB_2$ powder into the Al melt reduced the height that the dendritic and acicular grains grew from the Cu interface and promoted a transition to a cellular structure. Particles of $TiB_2$ within the melt serve as heterogeneous nucleation sites within the supercooled liquid ahead of a solidification front. While the exact chemistry and mechanisms at the nucleating surface has long been debated [86-88], grain refinement can be generally understood to occur when the critical undercooling for nucleation on an inoculant particle is less than the constitutional undercooling of the alloy (i.e., solid nucleation at a temperature closer to the alloy liquidus line) [89]. The nucleation of Al solid ahead of the dendrite tip leads to increased Cu solute concentration in the melt. Solute pileup at dendrite roots in Cu-Al melts has been previously observed in high-brilliance synchrotron X-radiation microscopy to cause remelting of the root and branch detachment, reducing their number density, and promoting a more acicular shape [90]. Fragmentation of dendrites during solidification also slows their growth velocity and promotes the observed transition to a cellular microstructure [91]. Increasing the $TiB_2$ concentration within a printed layer is expected to raise the nucleation rate within the undercooled zone and enhance the dendrite fragmentation mechanism, consistent with the observed trends in Figure 1. The complete suppression of columnar $Al_2Cu_3$ grains occurs when the $Al_2Cu$ phase forms before the acicular tip can grow. Here, the active compositional control of the melt pool at the Al/Cu interface while holding the laser, powder flow rate, toolpath, and conductive heat sinks constant is targeting the $Al_2Cu_3$ solidification interface velocity. As described by solidification theory, the growth rate at the tip of a dendrite or acicular grain is proportional to the concentration gradient at the phase front [92]. By increasing the concentration of Cu within the melt, the solute Al concentration inherently decreases and with it the dendrite/acicular tip growth velocity.

## 4. Summary

Al/Cu parts have been additively manufactured via LENS. The rapid quench rates in LENS produced a metastable phase mixture at the Al/Cu interface comprised of $Al_2Cu_3$ and $Al_2Cu$. The combination of layer-by-layer compositional control and melt pool kinetics allow for compositional gradients with step sizes finer than the powder diameter or dispersity. Inclusion of Cu into the melt pool from each underlying layer resulted in compositional grading over 4 mm between 67% and 99% Al with step functions between pure 0% Al, 40% Al, and 67% Al. However, grading within the Al/Cu bimetallic system is complicated by the preferential growth of dendritic structures and acicular grains at the LENS deposited interface. The inclusion of inoculants, specifically $TiB_2$, enabled root and branch detachment during dendritic growth in addition to slowing the solidification of $Al_2Cu_3$ such that $Al_2Cu$ formed earlier in the print. Further control of the melt pool chemistry through active compositional control in LENS accelerated the effects of the grain refiners to completely suppress the $Al_2Cu_3$ columnar growth down to its 1 μm thick seed layer.

**Acknowledgements**
The authors thank C. Profazi for EBSD sample preparation. This work was supported by the Sandia National Laboratory Directed Research and Development (LDRD) program. Sandia National Laboratories is a multimission laboratory managed and operated by National Technology and Engineering Solutions of Sandia, LLC., a wholly owned subsidiary of Honeywell International, Inc., for the U.S. Department of Energy's National Nuclear Security Administration under Contract No. DE-NA0003525. This work describes objective technical results and analysis. Any subjective views or opinions that might be expressed in the paper do not necessarily represent the views of the U.S. Department of Energy or the U.S. Government. On behalf of all authors, the corresponding author states that there is no conflict of interest.

# Supplementary material for:
# Refining microstructures in additively manufactured Al/Cu gradients through TiB2 inclusions

Michael J. Abere,[1] Hyein Choi,[1] Levi Van Bastian,[1] Luis Jauregui,[1] Tomas F. Babuska,[1] Mark. A Rodriguez,[1] Frank W. DelRio,[1] Shaun R. Whetten,[1] and Andrew B. Kustas[1]

[1]Sandia National Laboratories, Albuquerque, NM USA 87123

**S-1. SEM and EDS images of the deposition of pure Al on Cu**

SEM characterization of pure Al on Cu with insets of Al and Cu EDS images.

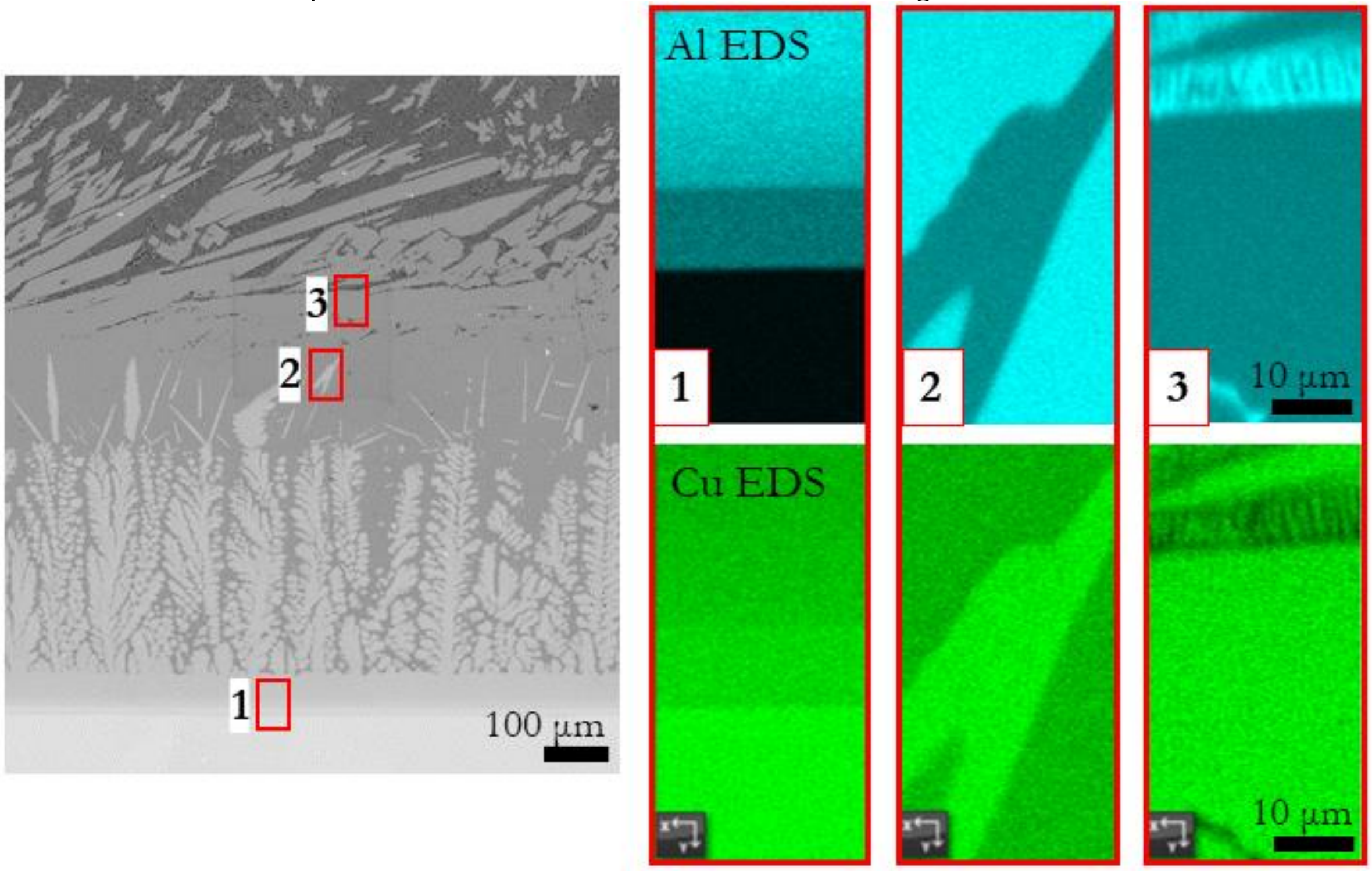

**Fig. S-1.** Al EDS and Cu EDS of each inset from Figure 2 are shown. Each inset from SEM correlates with an Al and Cu EDS corresponding to its numerical value. From these images, it is apparent that the cubic phase in the substrate of region 1 is Cu while in region 3 is Al.

**S-2. Band Contrast map of deposition of pure Al on Cu**

Band contrast map of first region from Figure 2 of SEM imaging of pure Al on Cu. From the map, an absence of well defined Kikuchi bands upon diffraction is depicted.

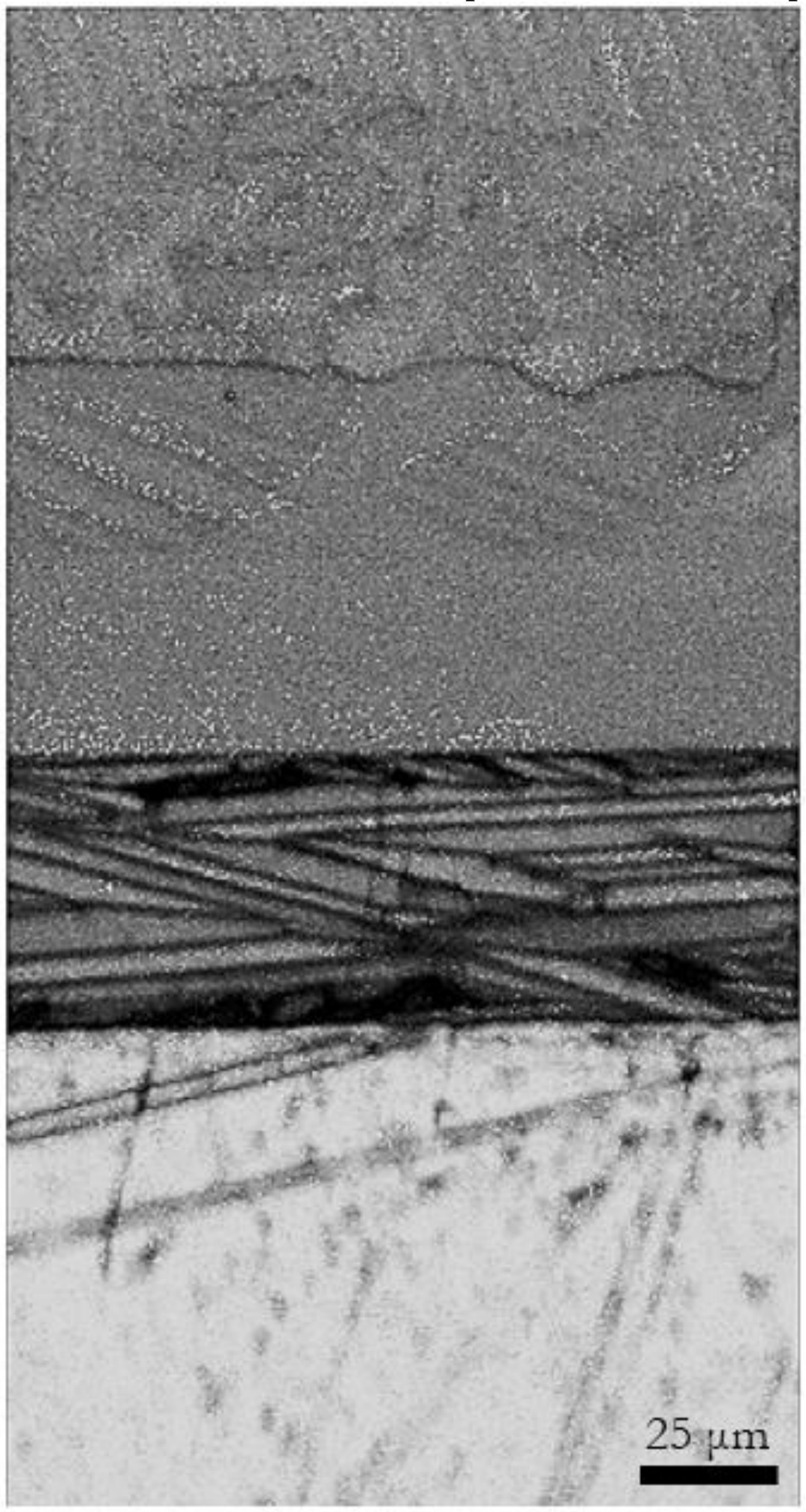

**Fig. S-2.** Band image of inset 1 from figure 2 is shown. The absence of well-defined Kikuchi bands suggests of a highly strained lattice.

## S-3. SEM and EBSD images of the Cu interface after deposition of 1% $TiB_2$

SEM imaging of 1% $TiB_2$ deposition with EBSD insets.

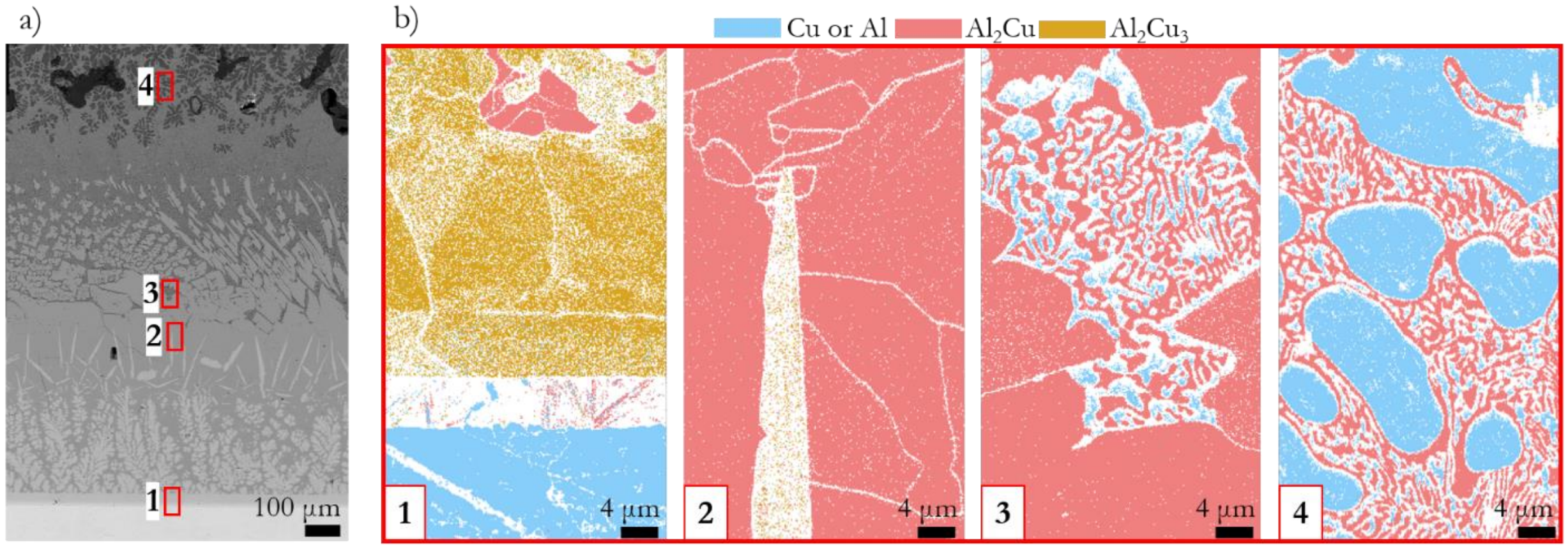

**Fig. S-3.** SEM image of the Cu interface after deposition of 1% $TiB_2$. (*a*) depicts larger area with rectangular boxes correlating to numbers 1 through 4. EBSD patterns taken from the inset locations of each numbered region can be seen in (*b*) with a legend of chemical compounds above.

**S-4. Full height of pure Al in SEM**

The full SEM image of the deposition of pure Al.

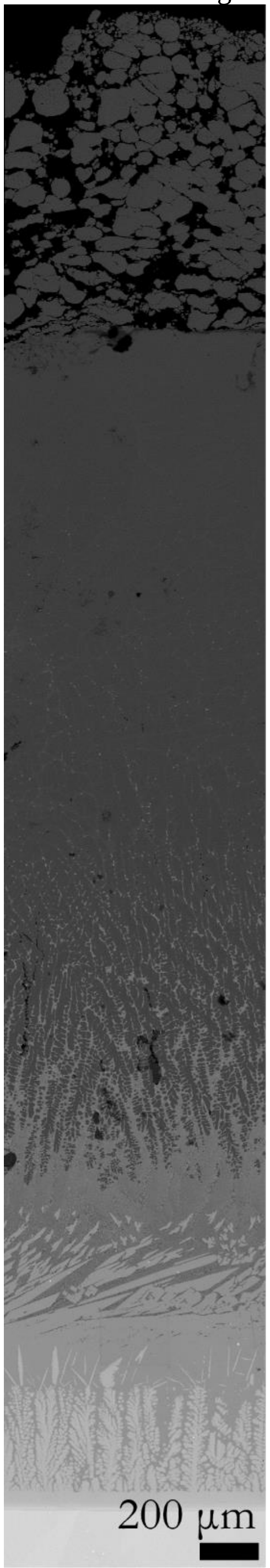

**Fig. S-4.** SEM of the complete height of the deposition of pure Al on Cu. Evidence of nucleation of hypoeutectic α-Al is seen from 1100 um above the Cu interface and persists to the top.

**S-5. SEM of pure Al with 2% $TiB_2$**

The SEM image of the deposition of Al with 2% $TiB_2$ from Figure 3. Inclusion of $TiB_2$ leads to nucleation of hypoeutectic Al earlier in the print compared to pure Al.

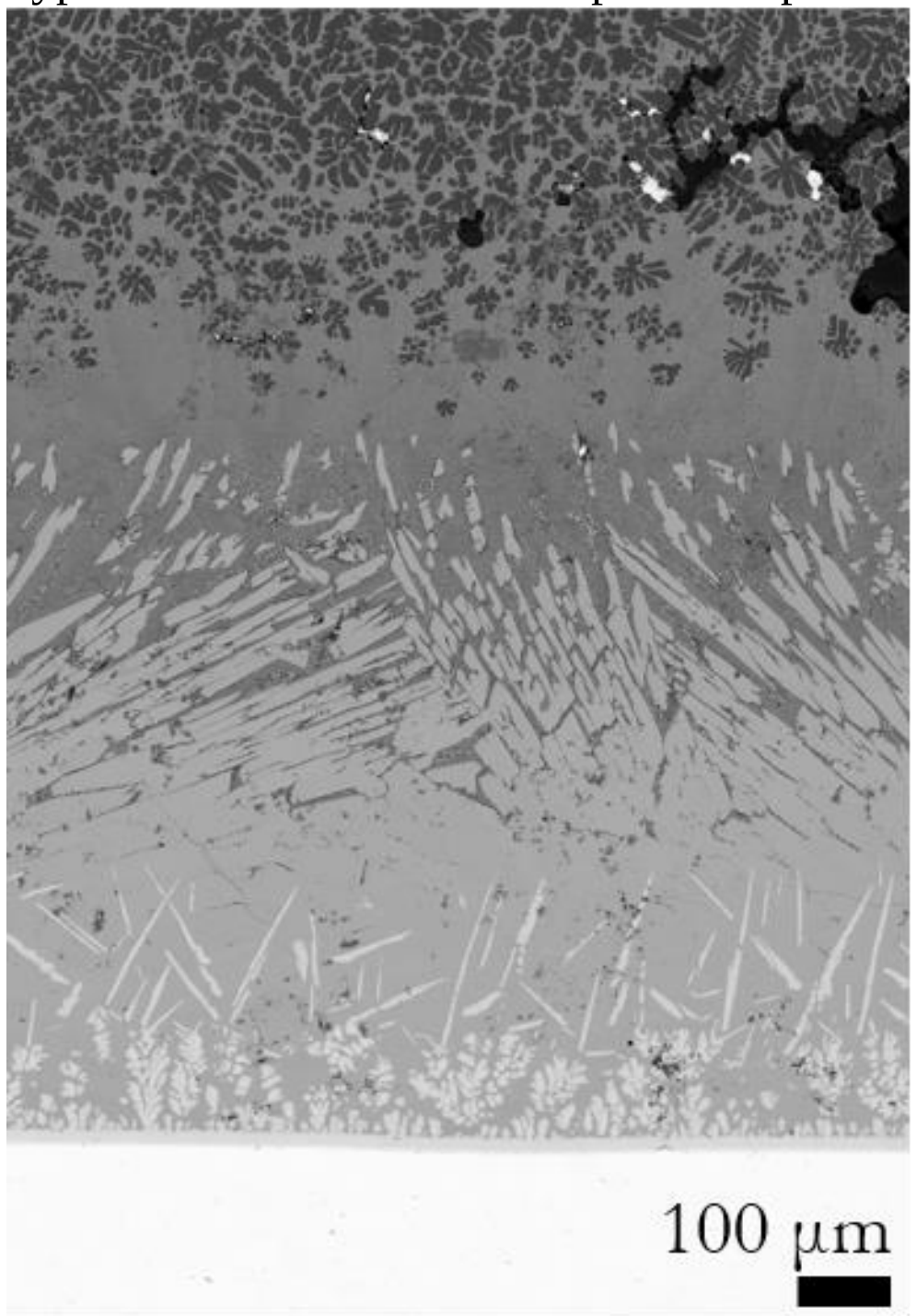

**Fig. S-5.** SEM of the complete height of the deposition of Al with 2%TiB2. At 800 µm above the Cu interface, hypoeutectic α-Al can be seen.

**S-6/7. Nanoindentation Hardness Measurements**

Additional nanoindentation results with corresponding optical images of the regions they were taken from are provided. The data shows that the phase specific hardness observed for pure Al printing does not change with the addition of $TiB_2$ inoculants.

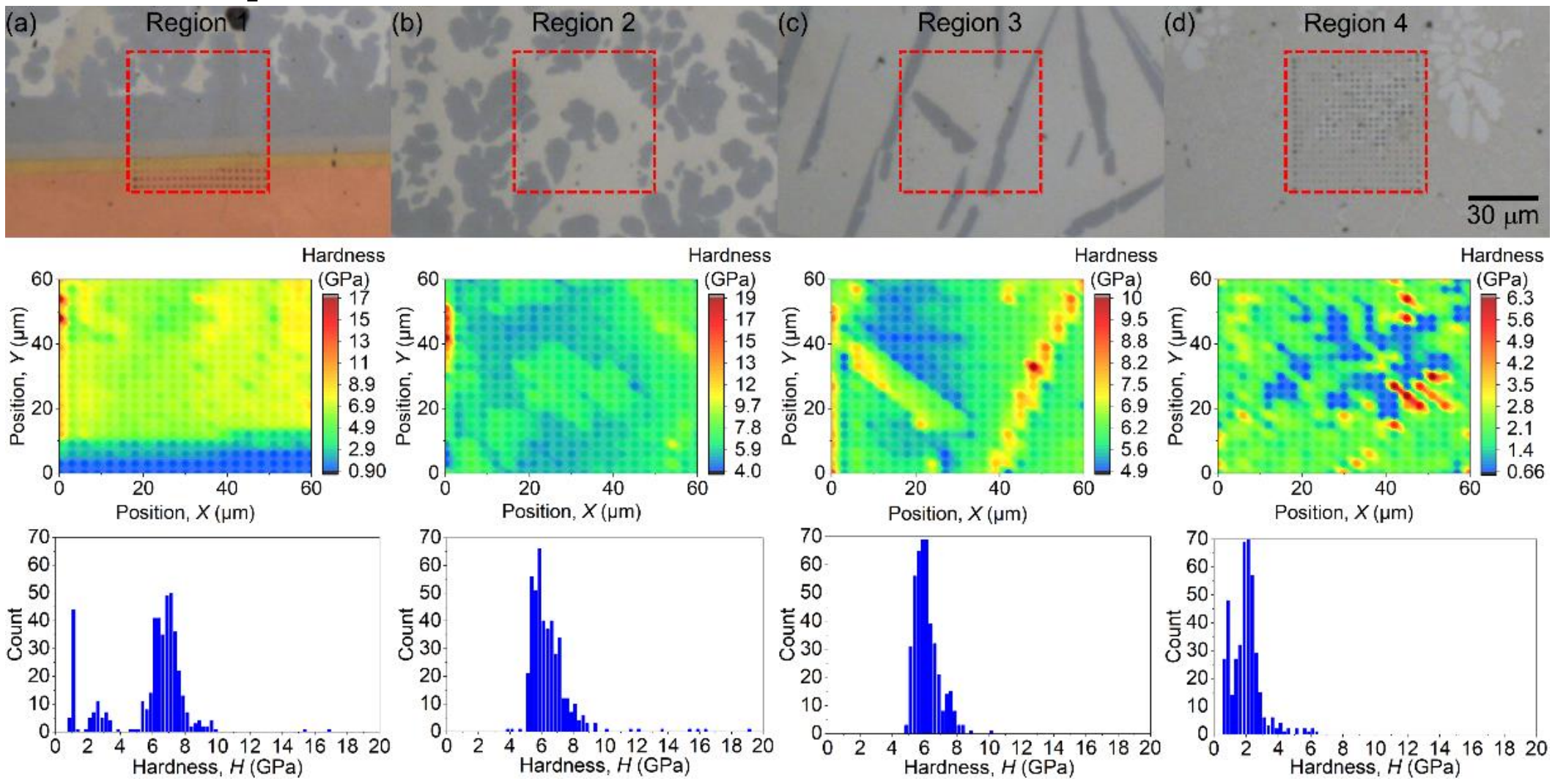

**Fig. S-6.** Nanoindentation hardness measurements of an Al print with 1 wt% TiB2 inoculant added on a Cu substrate showing local variation in different regions of the sample including (*a*) the Al/Cu interface, (*b*) the dendritic zone, (*c*) near acyclic grains, and (*d*) the Al-rich topmost area.

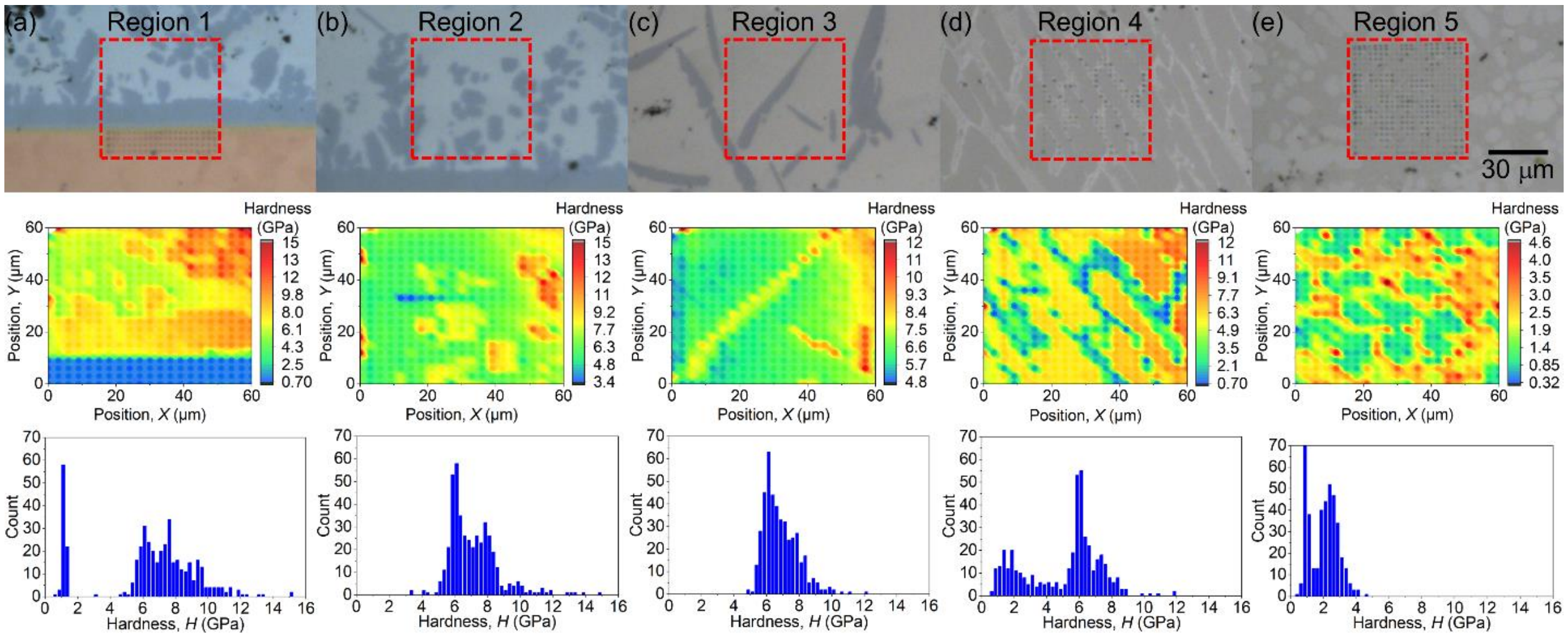

**Fig. S-7.** Nanoindentation hardness measurements of an Al print with 2 wt% TiB2 inoculant added on a Cu substrate showing local variation in different regions of the sample including (*a*) the Al/Cu interface, (*b*) the dendritic zone, (*c*) near acyclic grains, (*d*) and (*e*) two locations within the Al-rich topmost area.